\documentclass[11pt]{article}

\usepackage{amsmath}
\usepackage{amssymb}
\usepackage{amsfonts}

\usepackage{jheppub}

\usepackage{color}
\usepackage{axodraw4j}
\usepackage{pstricks}

\title{ An upper bound on the Kaon B-parameter and  Re$(\varepsilon_K)$}
 
\author{Jean-Marc GERARD}

\affiliation{Centre for Cosmology, Particle Physics and Phenomenology (CP3),\\
Universit\'e catholique de Louvain, B-1348, Louvain-la-Neuve, Belgium}

\emailAdd{jean-marc.gerard@uclouvain.be}

\abstract{New precise data in  B   physics and theoretical developments in  K  physics lead us to reconsider the weak $K^0\leftrightarrow \overline{K^0}$ transition from a large-N$_c$ viewpoint,  N$_c$ being the number of colors. In this framework, we infer an upper limit on $\hat{B}_K$ and the Kaon indirect CP violation.}

\begin{document}

\maketitle

\section{Introduction}

In the Standard Model (SM) for the electroweak interactions, the Cabibbo-Kobayashi-Maskawa (CKM) paradigm \cite{Cabibbo:1963yz,Kobayashi:1973fv} with its single phase uniquely relates $K^0$ to $B^0$ physics through time-dependent CP-symmetries defined by
\begin{equation}
\label{K1}
A^M_f (t) \equiv 
\frac{\mbox{Prob}\{\overline{M^0}(t) \to f_{\mbox{{\tiny CP}}}\} - \mbox{Prob}\{M^0(t) \to f_{\mbox{{\tiny CP}}}\} }{\mbox{Prob}\{\overline{M^0} (t) \to f_{\mbox{{\tiny CP}}}\} + \mbox{Prob}\{M^0(t) \to f_{\mbox{{\tiny CP}}}\}}
\end{equation}
if $f_{\mbox{{\tiny CP}}}$  is a common CP eigenstate.   As a matter of  fact, the most precise CP violation data available today \cite{Nakamura:2010zzi} can be expressed in the following way  
\begin{eqnarray}
&A^K_{\pi\pi} (t=0) = -2 \  \mbox{Re}(\varepsilon'_{K}) = - (5.0 \pm 0.8) 10^{-6}&
\label{K2a}\\
&A^B_{J/\Psi K} (t\approx 310^{-12} s) = \sin  2\beta =  0.678 \pm 0.025&
\label{K2b}\\
&A^K_{\pi\pi} (t=\infty) = +2 \  \mbox{Re}(\varepsilon_{K}) = + (3.32 \pm 0.04) 10^{-3}.&
 \label{K2c}
\end{eqnarray}
These asymmetries covering five orders of magnitude display the expected pattern for CP violation in pure decay, in decay-mixing interference and in pure mixing, respectively. Yet, nowadays they lead to some tension \cite{Lunghi:2008aa,Buras:2008nn,Buras:2009pj}. Indeed, if the angle $\beta$ is extracted from Eq.(\ref{K2b}) to fix the phase for the V$_{\mbox{{\scriptsize td}}}$ element of the unitary CKM matrix, then CP violation in the $K^0\to \pi\pi$ decays can be estimated in the SM and tends to be smaller than its experimental value. Such a statement mainly relies on quite precise estimates of the $|\Delta S| = 2$ hadronic matrix element from lattice QCD.   Conventionally normalized with respect to a naive vacuum insertion approximation (VIA) to be defined later on, these calculations with $n_f=2+1$ flavors have been reviewed and averaged very recently to give \cite{Colangelo:2010et}

\begin{equation}
\label{R3}
\hat{B}_K (\mbox{lattice}) = 0.724 \pm 0.024
\end{equation}
in the isospin limit. This dramatic reduction in errors for a numerical calculation of the scale invariant $\hat{B}_K$ parameter turns out to confirm an early analytical estimate simply performed in the large-$N_c$ limit ($N_c$ being the number of colors), namely   \cite{Gaiser:1980gx,Buras:1985yx}
\begin{equation}
\label{R4}
\hat{B}_K (N_c \to \infty) = \frac{3}{4}.
\end{equation}
It is thus quite tempting to extend this large-$N_c$ limit to the non-perturbative physics involved in the measured CP-asymmetry (\ref{K2c}). 

\paragraph{}In this paper we use the $1/N_c$ expansion to derive an upper bound on $\hat{B}_K$ and Re$(\varepsilon_{K})$, without input from lattice QCD. Section 2 gives the leading order, and Section 3 the sign of the next-to-leading corrections as well as arguments why the $1/N_c$ expansion for $\hat{B}_K$ is well-behaved. In Section 4, a phenomenological estimate of $\hat{B}_K$ from the observed $K^+\to \pi^+\pi^0$ decay is also given for consistency.

\section{The large-$N_c$ limit for Re$(\varepsilon_K)$}

At first order in CP violation, the asymmetry  (\ref{K2c}) is given by
\begin{equation}
\label{R5}
2 \ \mbox{Re}(\varepsilon_K) = \sin (2\phi_\varepsilon) \left\{\frac{\mbox{Im} M_{12}}{\Delta M} - \frac{\mbox{Im} \Gamma_{12}}{\Delta \Gamma}\right\} 
\end{equation}
with $\Delta M (\approx 2 \  \mbox{Re}M_{12})$, the $K_L-K_S$ mass splitting measured to be \cite{Nakamura:2010zzi} 
\begin{equation}
\label{R6}
\Delta M \equiv M_L - M_S = (3.483 \pm 0.006) 10^{-15} \ \mbox{GeV}
\end{equation}
and $\Delta \Gamma (\approx 2$ Re$\Gamma_{12})$, the corresponding decay width splitting that turns out to be of the same order since \cite{Nakamura:2010zzi}
\begin{equation}
\label{R7}
\phi_\varepsilon \equiv \mbox{tan}^{-1} \Bigl(\frac{2\Delta M}{-\Delta \Gamma}\Bigr) = (43.51 \pm 0.05)^\circ.
\end{equation}
Consequently, in Eq.(\ref{R5}), the only unknown quantities are the imaginary parts of the dispersive and absorptive contributions to the $|\Delta S| = 2$ weak transition:
\begin{eqnarray}
M_{12} &=& \langle \overline{K^0} |H_W^{\Delta S=2}| K^0 \rangle + \sum_n P \frac{\langle \overline{K^0}  |H_W^{\Delta S=1}| n \rangle \langle n |H_W^{\Delta S=1}| K^0 \rangle}{M_K-E_n},
\nonumber\\
\Gamma_{12} &=& 2\pi \sum_n \delta (M_K-E_n) \langle \overline{K^0}  |H_W^{\Delta S=1}| n \rangle \langle n |H_W^{\Delta S=1}| K^0 \rangle.
\label{R8}
\end{eqnarray}
In this general second-order perturbation formalism, the operator P projects out the principal part of $(M_K-E_n)^{-1}$  in $M_{12}$. Accordingly, the intermediate states $| n\rangle$ contributing to $M_{12}$ are virtual while the ones contributing to $\Gamma_{12}$ are physical states to which $K^0$ and $\overline{K^0}$ decay. So, both Im$M_{12}$ and Im$\Gamma_{12}$ are in principle affected by long-distance (LD) effects below the one GeV confining scale of QCD. Within the SM framework, these non-perturbative effects are generated by the so-called box and double-penguin diagrams \cite{Donoghue:1992dd,Branco:1999fs}. 
\\
In the large-$N_c$ limit \cite{'tHooft:1973jz}, internal quark loops are naturally suppressed and a full hadronization of these primary Feynman diagrams is performed with the help of planar gluons. At the leading ${\cal O} (N_c)$ level \cite{Bijnens:1990mz}, the weak $K^0\!\leftrightarrow\!\overline{K^0}$ transition allows virtual as well as physical $| (q\bar{q}) (q\bar{q})\rangle$ intermediate states (see the cut in Fig.1a) and thus contributes to both $M^{\mbox{{\tiny LD}}}_{12}$ and $\Gamma_{12}$, in a way compatible with Eq.(\ref{R7}).~As a consequence, in the limit of a large number of colors, Eq.(\ref{R5}) simply becomes 
\begin{equation}
\label{R9}
2 \ \mbox{Re}(\varepsilon_K) (N_c \to \infty) = \sin (2 \phi_\varepsilon) \ \frac{\mbox{Im}M^{\mbox{{\tiny SD}}}_{12}}{\Delta M}.
\end{equation}
In this expression,  $M^{\mbox{{\tiny SD}}}_{12}$ stands for the short-distance part of the $|\Delta S| = 2$ weak transition generated by the box-diagram with charm and top quarks propagating in the loop, once the standard phase convention for the CKM matrix (i.e., Im$V_{ud} V_{us}^\ast = 0)$ is adopted.   So the negative correction from $\phi_\varepsilon \neq \pi/4$ and CP violation in $K^0$ decays estimated in \cite{Buras:2010pza} for $|\varepsilon_K|$ does not affect Re$(\varepsilon_K)$, namely CP violation in the $K^0-\overline{K^0}$ mixing, in this limit. Taking into account the latest CKM input parameters and short-distance QCD corrections listed in \cite{Brod:2010mj}, we obtain 
\begin{equation}
\label{R10}
\mbox{Im}M^{\mbox{{\tiny SD}}}_{12} = \hat{B}_K (1.38 \pm 0.18) \  10^{-17} \ \mbox{GeV}.
\end{equation}
Here, the main uncertainty (about 10\%) comes from our limited knowledge of the c-to-b weak transition \cite{Nakamura:2010zzi},
\begin{equation}
\label{R11}
|V_{cb}| = (40.6 \pm 1.3) \ 10^{-3}.
\end{equation}
Indeed, Im$M^{\mbox{{\tiny SD}}}_{12}$ scales roughly as the fourth power of $|V_{cb}|$ if the 3-by-3 CKM matrix is taken unitary, as it should be in the SM. Using Eqs.(\ref{R4}), (\ref{R6}), (\ref{R7}) and (\ref{R10}) to evaluate Eq.(\ref{R9}), we conclude that the large-$N_c$ limit prediction for the $K^0-\overline{K^0}$ CP-asymmetry is
\begin{equation}
\label{R12}
2 \ \mbox{Re}(\varepsilon_K) (N_c \to \infty) = (2.96 \pm 0.38) \ 10^{-3}.
\end{equation}
This value is quite compatible with the experimental one given in Eq.(\ref{K2c}) but clearly favors a large value of $|V_{cb}|$, as extracted from inclusive decays in the Standard Model.

\begin{figure}[!h]
%%JaxoComment:
%%JaxoScale{1.0}
\SetScale{0.4}\setlength{\unitlength}{0.4pt}

\begin{center}
\fcolorbox{white}{white}{
  \begin{picture}(1122,280) (95,-27)
    \SetWidth{1.0}
    \SetColor{Black}
    \GOval(144,136)(48,16)(0){0.882}
    \Line(96,136)(128,136)
    \Line[arrow,arrowpos=0.5,arrowlength=5,arrowwidth=2,arrowinset=0.2](144,88)(224,88)
    \Line[arrow,arrowpos=0.5,arrowlength=5,arrowwidth=2,arrowinset=0.2](224,88)(224,184)
    \Line[arrow,arrowpos=0.5,arrowlength=5,arrowwidth=2,arrowinset=0.2](224,184)(144,184)
    \GOval(400,136)(48,16)(0){0.882}
    \Line[arrow,arrowpos=0.5,arrowlength=5,arrowwidth=2,arrowinset=0.2](400,184)(320,184)
    \Line[arrow,arrowpos=0.5,arrowlength=5,arrowwidth=2,arrowinset=0.2](320,184)(320,88)
    \Line[arrow,arrowpos=0.5,arrowlength=5,arrowwidth=2,arrowinset=0.2](320,88)(400,88)
    \Line(416,136)(448,136)
    \Photon(224,184)(320,184){7.5}{5}
    \Photon(224,88)(320,88){7.5}{5}
    \Text(176,200)[lb]{\normalsize{\Black{$s$}}}
    \DashBezier(344,208)(344,136)(272,184)(272,136){2}%JaxoID:GBez
    \DashBezier(272,136)(272,88)(200,136)(200,64){2}%JaxoID:GBez
    \Text(232,128)[lb]{\normalsize{\Black{$u$}}}
    \Text(296,128)[lb]{\normalsize{\Black{$u$}}}
    \Text(352,200)[lb]{\normalsize{\Black{$d$}}}
    \Text(176,48)[lb]{\normalsize{\Black{$d$}}}
    \Text(352,48)[lb]{\normalsize{\Black{$s$}}}
    \Text(264,200)[lb]{\normalsize{\Black{$W$}}}
    \Text(264,48)[lb]{\normalsize{\Black{$W$}}}
    \Text(256,-32)[lb]{\normalsize{\Black{$(a)$}}}
    \Line(496,136)(528,136)
    \GOval(544,136)(48,16)(0){0.882}
    \Line[arrow,arrowpos=0.5,arrowlength=5,arrowwidth=2,arrowinset=0.2](624,184)(544,184)
    \Line[arrow,arrowpos=0.5,arrowlength=5,arrowwidth=2,arrowinset=0.2](704,184)(624,184)
    \Line[arrow,arrowpos=0.5,arrowlength=5,arrowwidth=2,arrowinset=0.2](784,184)(704,184)
    \Photon(624,184)(624,88){7.5}{5}
    \Photon(704,184)(704,88){7.5}{5}
    \GOval(784,136)(48,16)(0){0.882}
    \Line(800,136)(832,136)
    \Line(880,136)(912,136)
    \GOval(928,136)(48,16)(0){0.882}
    \Line[arrow,arrowpos=0.5,arrowlength=5,arrowwidth=2,arrowinset=0.2](1008,184)(928,184)
    \Line[arrow,arrowpos=0.5,arrowlength=5,arrowwidth=2,arrowinset=0.2](1088,184)(1008,184)
    \Line[arrow,arrowpos=0.5,arrowlength=5,arrowwidth=2,arrowinset=0.2](1168,184)(1088,184)
    \Line[arrow,arrowpos=0.5,arrowlength=5,arrowwidth=2,arrowinset=0.2](544,88)(624,88)
    \Line[arrow,arrowpos=0.5,arrowlength=5,arrowwidth=2,arrowinset=0.2](624,88)(704,88)
    \Line[arrow,arrowpos=0.5,arrowlength=5,arrowwidth=2,arrowinset=0.2](704,88)(784,88)
    \PhotonArc[clock](1048,175)(41,167.32,12.68){7.5}{6.5}
    \Line[arrow,arrowpos=0.5,arrowlength=5,arrowwidth=2,arrowinset=0.2](928,88)(1008,88)
    \Line[arrow,arrowpos=0.5,arrowlength=5,arrowwidth=2,arrowinset=0.2](1008,88)(1088,88)
    \Line[arrow,arrowpos=0.5,arrowlength=5,arrowwidth=2,arrowinset=0.2](1088,88)(1168,88)
    \PhotonArc(1048,97)(41,-167.32,-12.68){7.5}{6.5}
    \GOval(1168,136)(48,16)(0){0.882}
    \Line(1184,136)(1216,136)
    \Text(576,200)[lb]{\normalsize{\Black{$s$}}}
    \Text(736,200)[lb]{\normalsize{\Black{$d$}}}
    \Text(656,200)[lb]{\normalsize{\Black{$u$}}}
    \Text(576,48)[lb]{\normalsize{\Black{$d$}}}
    \Text(656,48)[lb]{\normalsize{\Black{$u$}}}
    \Text(736,48)[lb]{\normalsize{\Black{$s$}}}
    \Text(584,128)[lb]{\normalsize{\Black{$W$}}}
    \Text(720,128)[lb]{\normalsize{\Black{$W$}}}
    \Text(1040,232)[lb]{\normalsize{\Black{$W$}}}
    \Text(1040,16)[lb]{\normalsize{\Black{$W$}}}
    \Text(960,200)[lb]{\normalsize{\Black{$s$}}}
    \Text(960,48)[lb]{\normalsize{\Black{$d$}}}
    \Text(1120,200)[lb]{\normalsize{\Black{$d$}}}
    \Text(1120,48)[lb]{\normalsize{\Black{$s$}}}
    \Text(648,-32)[lb]{\normalsize{\Black{$(b)$}}}
    \Text(1032,-32)[lb]{\normalsize{\Black{$(c)$}}}
    \DashBezier(1112,208)(1112,120)(984,152)(984,64){2}%JaxoID:GBez
    \Bezier(1008,136)(1008,160)(1088,160)(1088,136)\Line[arrow,arrowpos=0.5,arrowlength=5,arrowwidth=2,arrowinset=0.2](1047.4,153.998)(1048,154)%JaxoID:FBez[arrow,arrowpos=0.5,arrowlength=5,arrowwidth=2,arrowinset=0.2]
    \Bezier(1088,136)(1088,112)(1008,112)(1008,136)\Line[arrow,arrowpos=0.5,arrowlength=5,arrowwidth=2,arrowinset=0.2](1048.6,118.002)(1048,118)%JaxoID:FBez[arrow,arrowpos=0.5,arrowlength=5,arrowwidth=2,arrowinset=0.2]
  \end{picture}
}
\end{center}
\caption{a) $\mathcal{O}\left(N_c\right)$, b) $\mathcal{O}\left(1\right)$ and c) $\mathcal{O}\left(1/N_c\right)$ non-perturbative contributions to the $K^0\leftrightarrow\bar{K}^0$ transition. Dotted cuts identify the physical intermediate states for $\Gamma_{12}$.}
\label{Fig:transi}
\end{figure}
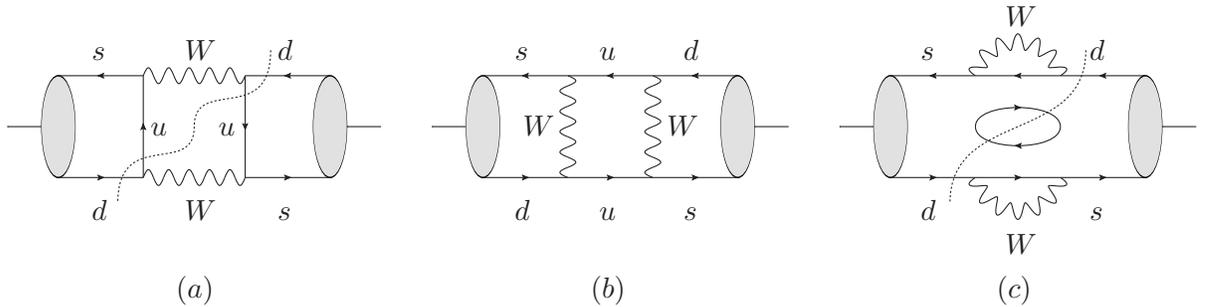
 
\section{The $1/N_c$ corrections to Re$(\varepsilon_K)$ and $\hat{B}_K$}

Before excluding any possible tension between Re$(\varepsilon_K)$ and $\sin 2\beta$ in the SM, we still have to estimate the size of the $1/N_c$ corrections.  At the next-to-leading ${\cal O} (1)$ level, $M^{\mbox{{\tiny LD}}}_{12}$ stays real (see $V_{ud}V_{us}^*$ in Fig.1b) and Im$\Gamma_{12}$ is still zero (see the cut in Fig.1c). So, at this level the LD contributions considered in \cite{Buras:2010pza} are $1/N_c^2$-suppressed and only the $1/N_c$ correction to the B-parameter of the  dominant $\Delta S = 2$ hadronic matrix element
\begin{equation}
\label{R13}
\langle \overline{K^0} | [\bar{s} \gamma_\mu (1-\gamma^5) d] \otimes [\bar{s} \gamma^\mu (1-\gamma^5)d] | K^0 \rangle \equiv 2 \left(1+\frac{1}{N_c}\right) B_K f^2_K M^2_K
\end{equation}
in $M^{\mbox{{\tiny SD}}}_{12}$ really matters. One easily checks that the VIA corresponds to $B_K = 1$, as it should be \cite{Gaillard:1974hs}, since any (V-A)(V-A) four-quark operator obeys the standard Fierz relation
\begin{equation}
\label{R14}
[\bar{q}_i \gamma_\mu (1-\gamma^5) q^j] \otimes [\bar{q}_k \gamma^\mu (1-\gamma^5) q^l] = [\bar{q}_i \gamma_\mu (1-\gamma^5) q^l] \otimes [\bar{q}_k \gamma^\mu (1-\gamma^5)q^j].
\end{equation}
In the isospin limit taken hereafter, the measured decay constant $f_K \  (\approx 156$ MeV) of the Kaons is defined by the $K$-to-vacuum matrix elements for color-singlet left-handed currents
\begin{equation}
\label{R15}
\langle 0 | [\bar{s} \gamma^\mu (1-\gamma^5)d] | K^0 \rangle = \langle 0 | [\bar{s} \gamma^\mu (1-\gamma^5)u] | K^+ \rangle \equiv i f_K p^\mu.
\end{equation}
So, the explicit $1/N_c$ factor in Eq.(\ref{R13}) simply stems from the projection of the currents on the {\it color-singlet} $K^0$ state once the vacuum state has been inserted in all possible ways with the help of the Fierz relation (\ref{R14}). 
\\
Let us introduce a fictitious color-singlet $X^0$ boson exchange between the two $|\Delta S| = 1$ left-handed currents in Eq.(\ref{R13}). In fact, such a trick has already been used in the past to keep track of the loop-momentum flow when matching the SD operator evolution in QCD with the LD one estimated either in a Nambu-Jona-Lasinio model \cite{Bijnens:1994cx} or in a non-linear $\sigma$-model \cite{Fatelo:1994qh}. Here the X-boson propagator is not cut-off, but the full leading ${\cal O}(N_c)$ and next-to-leading ${\cal O}(1)$ contributions to $\hat{B}_K$ can be displayed as in Fig.2 without any reference to an effective model below one GeV. In particular, the Fierz relation (\ref{R14}) is partially at work between Fig.2a and Fig.2b. Following the standard Feynman rule that tells us to take the trace {\it and} to multiply by a factor (-1) for each closed fermion loop, we thus expect a negative $1/N_c$ correction for the Kaon $B$-parameter, namely
\begin{equation}
\label{R16}
 \hat{B}_K  (1/N_c) < \frac{3}{4}.
\end{equation}

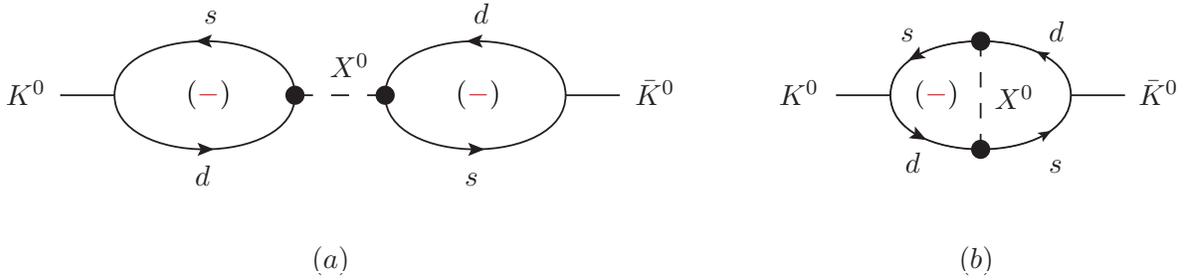
\begin{figure}[!h]
%%JaxoComment:
%%JaxoScale{1.0}
\SetScale{0.85}\setlength{\unitlength}{0.85pt}

\begin{center}
\fcolorbox{white}{white}{
  \begin{picture}(536,128) (19,-43)
    \SetWidth{0.1}
    \SetColor{Black}
    \Bezier(408,32)(408,64)(488,64)(488,32)\Line[arrow,arrowpos=0.75,arrowlength=3.5,arrowwidth=1.4,arrowinset=0.2,flip](475.047,50.238)(475.5,50)%JaxoID:FBez[arrow,arrowpos=0.75,arrowlength=3.5,arrowwidth=1.4,arrowinset=0.2,flip]
    \Bezier(488,32)(488,0)(408,0)(408,32)\Line[arrow,arrowpos=0.24,arrowlength=3.5,arrowwidth=1.4,arrowinset=0.2,flip](476.822,14.742)(476.388,14.49)%JaxoID:FBez[arrow,arrowpos=0.24,arrowlength=3.5,arrowwidth=1.4,arrowinset=0.2,flip]
    \SetWidth{0.8}
    \Bezier(488,32)(488,0)(408,0)(408,32)\Line[arrow,arrowpos=0.75,arrowlength=5,arrowwidth=2,arrowinset=0.2,flip](420.953,13.762)(420.5,14)%JaxoID:FBez[arrow,arrowpos=0.75,arrowlength=5,arrowwidth=2,arrowinset=0.2,flip]
    \Line(488,32)(512,32)
    \Line(384,32)(408,32)
    \Line(40,32)(64,32)
    \Line[dash,dashsize=8](144,32)(184,32)
    \Bezier(144,32)(144,0)(64,0)(64,32)\Line[arrow,arrowpos=0.5,arrowlength=5,arrowwidth=2,arrowinset=0.2,flip](104.6,8.002)(104,8)%JaxoID:FBez[arrow,arrowpos=0.5,arrowlength=5,arrowwidth=2,arrowinset=0.2,flip]
    \Bezier(64,32)(64,64)(144,64)(144,32)\Line[arrow,arrowpos=0.5,arrowlength=5,arrowwidth=2,arrowinset=0.2,flip](103.4,55.998)(104,56)%JaxoID:FBez[arrow,arrowpos=0.5,arrowlength=5,arrowwidth=2,arrowinset=0.2,flip]
    \Bezier(184,32)(184,64)(264,64)(264,32)\Line[arrow,arrowpos=0.5,arrowlength=5,arrowwidth=2,arrowinset=0.2,flip](223.4,55.998)(224,56)%JaxoID:FBez[arrow,arrowpos=0.5,arrowlength=5,arrowwidth=2,arrowinset=0.2,flip]
    \Bezier(264,32)(264,0)(184,0)(184,32)\Line[arrow,arrowpos=0.5,arrowlength=5,arrowwidth=2,arrowinset=0.2,flip](224.6,8.002)(224,8)%JaxoID:FBez[arrow,arrowpos=0.5,arrowlength=5,arrowwidth=2,arrowinset=0.2,flip]
    \Bezier(408,32)(408,64)(488,64)(488,32)\Line[arrow,arrowpos=0.24,arrowlength=5,arrowwidth=2,arrowinset=0.2,flip](419.178,49.258)(419.612,49.51)%JaxoID:FBez[arrow,arrowpos=0.24,arrowlength=5,arrowwidth=2,arrowinset=0.2,flip]
    \Line(264,32)(288,32)
    \Line[dash,dashsize=8](448,56)(448,8)
    \Text(16,28)[lb]{\normalsize{\Black{$K^0$}}}
    \Text(96,26)[lb]{\normalsize{$(\Red{-})$}}
    \Text(104,64)[lb]{\normalsize{\Black{$s$}}}
    \Text(100,-8)[lb]{\normalsize{\Black{$d$}}}
    \Text(160,40)[lb]{\normalsize{\Black{$X^0$}}}
    \Text(216,26)[lb]{\normalsize{$(\Red{-})$}}
    \Text(224,64)[lb]{\normalsize{\Black{$d$}}}
    \Text(220,-8)[lb]{\normalsize{\Black{$s$}}}
    \Text(296,28)[lb]{\normalsize{\Black{$\bar{K}^0$}}}
    \SetWidth{0.0}
    \Vertex(144,32){4.243}
    \Vertex(184,32){4.243}
    \Vertex(448,56){4.243}
    \Vertex(448,8){4.243}
    \Text(152,-48)[lb]{\normalsize{\Black{$(a)$}}}
    \Text(440,-48)[lb]{\normalsize{\Black{$(b)$}}}
    \Text(360,28)[lb]{\normalsize{\Black{$K^0$}}}
    \Text(414,56)[lb]{\normalsize{\Black{$s$}}}
    \Text(480,56)[lb]{\normalsize{\Black{$d$}}}
    \Text(416,-2)[lb]{\normalsize{\Black{$d$}}}
    \Text(480,-2)[lb]{\normalsize{\Black{$s$}}}
    \Text(520,28)[lb]{\normalsize{\Black{$\bar{K}^0$}}}
    \Text(456,26)[lb]{\normalsize{\Black{$X^0$}}}
    \Text(420,26)[lb]{\normalsize{$(\Red{-})$}}
  \end{picture}
}
\end{center}
\caption{a) $\mathcal{O}\left(N_c\right)$ and b) $\mathcal{O}\left(1\right)$ effective contributions to the $\hat{B}_K$ parameter. From now on, sum over planar gluons inside the ellipses is understood. Minus signs in parenthesis correspond to the Feynman factor for fermion loops and lead to a negative $1/N_c$ correction.}
\label{Fig:BK}
\end{figure}

Consequently, Eq.(\ref{R12}) should also be taken as an upper limit on Re$(\varepsilon_K)$.~Of course, these bounds hold true if such a $1/N_c$ expansion in the $K^0\leftrightarrow\overline{K^0}$ transition makes sense.   In this connection, earlier works inspired by the $\pi^+-\pi^0$ electromagnetic mass difference \cite{Bardeen:1988zw} have put forward a relatively smooth $1/N_c$ matching between the perturbative QCD theory and a truncated effective theory for the light pseudo-scalar \cite{Bardeen:1987vg} and vector \cite{Gerard:1990dx} mesons.   The central value quoted for $\hat{B}_K$ in these works were respectively 0.66 and 0.70, but with too large uncertainties to claim to be below $3/4$.        Later on, the inclusion of an extended Nambu-Jona-Lasinio model to improve the $1/N_c$ matching around the critical one GeV scale supported the range $0.60 < \hat{B}_K < 0.80$ \cite{Bijnens:1995br}.   As we now know,  the latest lattice results averaged in Eq.(\ref{R3}) suggest   very small $1/N_c$ corrections to $\hat{B}_K=0.75$.

The expected negative sign of the $1/N_c$ correction to the $\hat{B}_K$ parameter can be understood in the following pictorial way. If we substitute the $W^+$ gauge boson for the $X^0$ boson in Fig.2, we then obtain the two diagrams in Fig.3 that generate the full  $|\Delta S| = 1$ $K^+ \to \pi^+ \pi^0$ weak decay amplitude at the next-to-leading order, and in the isospin limit. Indeed, the third so-called penguin diagram added in Fig.4 only enhances the  $\Delta I = 1/2$ component of the $K^0 \to \pi \pi$ amplitudes and has no counterpart for the effective $|\Delta S| = 2$ transition in Fig.2.  In other words, the $1/N_c$ corrections for $\hat{B}_K$ and the $\Delta I =  3/2$ $K^+ \to \pi^+\pi^0$ amplitude are topologically equivalent and, consequently, negative since the $\Delta I =  3/2$ transition comes out too large at the leading order. 

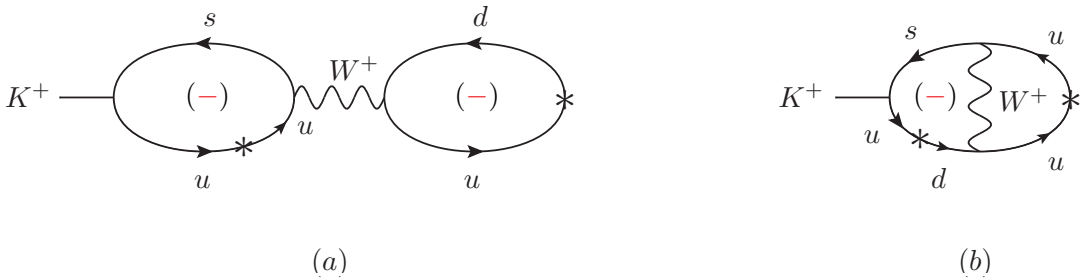
\begin{figure}[!h]
%%JaxoComment:
%%JaxoScale{1.0}
\SetScale{0.85}\setlength{\unitlength}{0.85pt}

\begin{center}
\fcolorbox{white}{white}{
  \begin{picture}(501,128) (19,-43)
    \SetWidth{0.1}
    \SetColor{Black}
    \Bezier(488,32)(488,0)(408,0)(408,32)\Line[arrow,arrowpos=0.64,arrowlength=3.5,arrowwidth=1.4,arrowinset=0.2,flip](432.194,9.75)(431.639,9.882)%JaxoID:FBez[arrow,arrowpos=0.64,arrowlength=3.5,arrowwidth=1.4,arrowinset=0.2,flip]
    \SetWidth{0.8}
    \Line(384,32)(408,32)
    \Bezier(408,32)(408,64)(488,64)(488,32)\Line[arrow,arrowpos=0.24,arrowlength=5,arrowwidth=2,arrowinset=0.2,flip](419.178,49.258)(419.612,49.51)%JaxoID:FBez[arrow,arrowpos=0.24,arrowlength=5,arrowwidth=2,arrowinset=0.2,flip]
    \Bezier(64,32)(64,64)(144,64)(144,32)\Line[arrow,arrowpos=0.5,arrowlength=5,arrowwidth=2,arrowinset=0.2,flip](103.4,55.998)(104,56)%JaxoID:FBez[arrow,arrowpos=0.5,arrowlength=5,arrowwidth=2,arrowinset=0.2,flip]
    \SetWidth{0.1}
    \Bezier(408,32)(408,64)(488,64)(488,32)\Line[arrow,arrowpos=0.75,arrowlength=3.5,arrowwidth=1.4,arrowinset=0.2,flip](475.047,50.238)(475.5,50)%JaxoID:FBez[arrow,arrowpos=0.75,arrowlength=3.5,arrowwidth=1.4,arrowinset=0.2,flip]
    \Bezier(488,32)(488,0)(408,0)(408,32)\Line[arrow,arrowpos=0.24,arrowlength=3.5,arrowwidth=1.4,arrowinset=0.2,flip](476.822,14.742)(476.388,14.49)%JaxoID:FBez[arrow,arrowpos=0.24,arrowlength=3.5,arrowwidth=1.4,arrowinset=0.2,flip]
    \SetWidth{0.8}
    \Bezier(488,32)(488,0)(408,0)(408,32)\Line[arrow,arrowpos=0.87,arrowlength=5,arrowwidth=2,arrowinset=0.2,flip](411.98,20.79)(411.704,21.142)%JaxoID:FBez[arrow,arrowpos=0.87,arrowlength=5,arrowwidth=2,arrowinset=0.2,flip]
    \Line(40,32)(64,32)
    \Bezier(144,32)(144,0)(64,0)(64,32)\Line[arrow,arrowpos=0.5,arrowlength=5,arrowwidth=2,arrowinset=0.2,flip](104.6,8.002)(104,8)%JaxoID:FBez[arrow,arrowpos=0.5,arrowlength=5,arrowwidth=2,arrowinset=0.2,flip]
    \Bezier(184,32)(184,64)(264,64)(264,32)\Line[arrow,arrowpos=0.5,arrowlength=5,arrowwidth=2,arrowinset=0.2,flip](223.4,55.998)(224,56)%JaxoID:FBez[arrow,arrowpos=0.5,arrowlength=5,arrowwidth=2,arrowinset=0.2,flip]
    \Bezier(264,32)(264,0)(184,0)(184,32)\Line[arrow,arrowpos=0.5,arrowlength=5,arrowwidth=2,arrowinset=0.2,flip](224.6,8.002)(224,8)%JaxoID:FBez[arrow,arrowpos=0.5,arrowlength=5,arrowwidth=2,arrowinset=0.2,flip]
    \Text(16,28)[lb]{\normalsize{\Black{$K^+$}}}
    \Text(96,26)[lb]{\normalsize{$(\Red{-})$}}
    \Text(104,64)[lb]{\normalsize{\Black{$s$}}}
    \Text(100,-8)[lb]{\normalsize{\Black{$u$}}}
    \Text(160,40)[lb]{\normalsize{\Black{$W^+$}}}
    \Text(216,26)[lb]{\normalsize{$(\Red{-})$}}
    \Text(224,64)[lb]{\normalsize{\Black{$d$}}}
    \Text(220,-8)[lb]{\normalsize{\Black{$u$}}}
    \Text(152,-48)[lb]{\normalsize{\Black{$(a)$}}}
    \Text(440,-48)[lb]{\normalsize{\Black{$(b)$}}}
    \Text(360,28)[lb]{\normalsize{\Black{$K^+$}}}
    \Text(416,58)[lb]{\normalsize{\Black{$s$}}}
    \Text(480,56)[lb]{\normalsize{\Black{$u$}}}
    \Text(428,-8)[lb]{\normalsize{\Black{$d$}}}
    \Text(480,0)[lb]{\normalsize{\Black{$u$}}}
    \Text(458,26)[lb]{\normalsize{\Black{$W^+$}}}
    \Text(420,26)[lb]{\normalsize{$(\Red{-})$}}
    \Photon(144,32)(184,32){5}{3}
    \Text(260,26)[lb]{\LARGE{\Black{$*$}}}
    \Text(117,5)[lb]{\LARGE{\Black{$*$}}}
    \SetWidth{0.1}
    \Bezier(144,32)(144,0)(64,0)(64,32)\Line[arrow,arrowpos=0.165,arrowlength=3.5,arrowwidth=1.4,arrowinset=0.2,flip](138.511,19.098)(138.185,18.774)%JaxoID:FBez[arrow,arrowpos=0.165,arrowlength=3.5,arrowwidth=1.4,arrowinset=0.2,flip]
    \Text(398,12)[lb]{\normalsize{\Black{$u$}}}
    \Text(418,9)[lb]{\LARGE{\Black{$*$}}}
    \Text(485,26)[lb]{\LARGE{\Black{$*$}}}
    \SetWidth{0.8}
    \Photon(448,56)(448,8){5}{3}
    \Text(146,16)[lb]{\normalsize{\Black{$u$}}}
  \end{picture}
}
\end{center}
\caption{a) $\mathcal{O}\left(N_c^{1/2}\right)$ and b) $\mathcal{O}\left(N_c^{-1/2}\right)$ contributions to the $K^+\to\pi^+\pi^0$ amplitude. Stars denote projections on the $\pi$ final states in the isospin limit. The minus signs correspond again to the Feynman factor for fermion loops and lead to a $\Delta I=3/2$ suppression.}\label{Fig:Kp}
\end{figure}

\begin{figure}[!h]
%%JaxoComment:
%%JaxoScale{1.0}
\SetScale{0.8}\setlength{\unitlength}{0.8pt}

\begin{center}
\fcolorbox{white}{white}{
  \begin{picture}(568,128) (25,-43)
    \SetWidth{0.1}
    \SetColor{Black}
    \Bezier(488,32)(488,64)(568,64)(568,32)\Line[arrow,arrowpos=0.75,arrowlength=3.5,arrowwidth=1.4,arrowinset=0.2,flip](555.047,50.238)(555.5,50)%JaxoID:FBez[arrow,arrowpos=0.75,arrowlength=3.5,arrowwidth=1.4,arrowinset=0.2,flip]
    \Bezier(488,32)(488,0)(568,0)(568,32)\Line[arrow,arrowpos=0.75,arrowlength=3.5,arrowwidth=1.4,arrowinset=0.2](555.047,13.762)(555.5,14)%JaxoID:FBez[arrow,arrowpos=0.75,arrowlength=3.5,arrowwidth=1.4,arrowinset=0.2]
    \Bezier(416,32)(416,0)(336,0)(336,32)\Line[arrow,arrowpos=0.24,arrowlength=3.5,arrowwidth=1.4,arrowinset=0.2,flip](404.822,14.742)(404.388,14.49)%JaxoID:FBez[arrow,arrowpos=0.24,arrowlength=3.5,arrowwidth=1.4,arrowinset=0.2,flip]
    \SetWidth{0.8}
    \Bezier(416,32)(416,0)(336,0)(336,32)\Line[arrow,arrowpos=0.87,arrowlength=5,arrowwidth=2,arrowinset=0.2,flip](339.98,20.79)(339.704,21.142)%JaxoID:FBez[arrow,arrowpos=0.87,arrowlength=5,arrowwidth=2,arrowinset=0.2,flip]
    \SetWidth{0.1}
    \Bezier(416,32)(416,0)(336,0)(336,32)\Line[arrow,arrowpos=0.64,arrowlength=3.5,arrowwidth=1.4,arrowinset=0.2,flip](360.194,9.75)(359.639,9.882)%JaxoID:FBez[arrow,arrowpos=0.64,arrowlength=3.5,arrowwidth=1.4,arrowinset=0.2,flip]
    \SetWidth{0.8}
    \Line(316,32)(336,32)
    \Bezier(336,32)(336,64)(416,64)(416,32)\Line[arrow,arrowpos=0.24,arrowlength=5,arrowwidth=2,arrowinset=0.2,flip](347.178,49.258)(347.612,49.51)%JaxoID:FBez[arrow,arrowpos=0.24,arrowlength=5,arrowwidth=2,arrowinset=0.2,flip]
    \Bezier(488,32)(488,64)(568,64)(568,32)\Line[arrow,arrowpos=0.385,arrowlength=5,arrowwidth=2,arrowinset=0.2,flip](513.876,54.618)(514.443,54.73)%JaxoID:FBez[arrow,arrowpos=0.385,arrowlength=5,arrowwidth=2,arrowinset=0.2,flip]
    \Line(468,32)(488,32)
    \Bezier(568,32)(568,0)(488,0)(488,32)\Line[arrow,arrowpos=0.87,arrowlength=5,arrowwidth=2,arrowinset=0.2,flip](491.98,20.79)(491.704,21.142)%JaxoID:FBez[arrow,arrowpos=0.87,arrowlength=5,arrowwidth=2,arrowinset=0.2,flip]
    \SetWidth{0.1}
    \Bezier(568,32)(568,0)(488,0)(488,32)\Line[arrow,arrowpos=0.49,arrowlength=3.5,arrowwidth=1.4,arrowinset=0.2,flip](529.799,8.022)(529.2,8.01)%JaxoID:FBez[arrow,arrowpos=0.49,arrowlength=3.5,arrowwidth=1.4,arrowinset=0.2,flip]
    \Bezier(488,32)(488,0)(568,0)(568,32)\Line[arrow,arrowpos=0.33,arrowlength=3.5,arrowwidth=1.4,arrowinset=0.2](507.857,10.94)(508.386,10.774)%JaxoID:FBez[arrow,arrowpos=0.33,arrowlength=3.5,arrowwidth=1.4,arrowinset=0.2]
    \Bezier(336,32)(336,64)(416,64)(416,32)\Line[arrow,arrowpos=0.75,arrowlength=3.5,arrowwidth=1.4,arrowinset=0.2,flip](403.047,50.238)(403.5,50)%JaxoID:FBez[arrow,arrowpos=0.75,arrowlength=3.5,arrowwidth=1.4,arrowinset=0.2,flip]
    \SetWidth{0.8}
    \Bezier(64,32)(64,64)(144,64)(144,32)\Line[arrow,arrowpos=0.5,arrowlength=5,arrowwidth=2,arrowinset=0.2,flip](103.4,55.998)(104,56)%JaxoID:FBez[arrow,arrowpos=0.5,arrowlength=5,arrowwidth=2,arrowinset=0.2,flip]
    \Line(44,32)(64,32)
    \Bezier(144,32)(144,0)(64,0)(64,32)\Line[arrow,arrowpos=0.5,arrowlength=5,arrowwidth=2,arrowinset=0.2,flip](104.6,8.002)(104,8)%JaxoID:FBez[arrow,arrowpos=0.5,arrowlength=5,arrowwidth=2,arrowinset=0.2,flip]
    \Bezier(184,32)(184,64)(264,64)(264,32)\Line[arrow,arrowpos=0.5,arrowlength=5,arrowwidth=2,arrowinset=0.2,flip](223.4,55.998)(224,56)%JaxoID:FBez[arrow,arrowpos=0.5,arrowlength=5,arrowwidth=2,arrowinset=0.2,flip]
    \Bezier(264,32)(264,0)(184,0)(184,32)\Line[arrow,arrowpos=0.5,arrowlength=5,arrowwidth=2,arrowinset=0.2,flip](224.6,8.002)(224,8)%JaxoID:FBez[arrow,arrowpos=0.5,arrowlength=5,arrowwidth=2,arrowinset=0.2,flip]
    \Text(22,28)[lb]{\normalsize{\Black{$K^0$}}}
    \Text(94,26)[lb]{\normalsize{$(\Red{-})$}}
    \Text(104,64)[lb]{\normalsize{\Black{$s$}}}
    \Text(100,-8)[lb]{\normalsize{\Black{$d$}}}
    \Text(160,40)[lb]{\normalsize{\Black{$W^+$}}}
    \Text(216,26)[lb]{\normalsize{$(\Red{-})$}}
    \Text(224,64)[lb]{\normalsize{\Black{$d$}}}
    \Text(220,-8)[lb]{\normalsize{\Black{$u$}}}
    \Text(152,-48)[lb]{\normalsize{\Black{$(a)$}}}
    \Text(368,-48)[lb]{\normalsize{\Black{$(b)$}}}
    \Text(294,28)[lb]{\normalsize{\Black{$K^0$}}}
    \Text(344,56)[lb]{\normalsize{\Black{$s$}}}
    \Text(408,56)[lb]{\normalsize{\Black{$u$}}}
    \Text(356,-8)[lb]{\normalsize{\Black{$d$}}}
    \Text(408,2)[lb]{\normalsize{\Black{$u$}}}
    \Text(384,26)[lb]{\normalsize{\Black{$W^+$}}}
    \Text(348,26)[lb]{\normalsize{$(\Red{-})$}}
    \Photon(144,32)(184,32){5}{3}
    \Text(260,26)[lb]{\LARGE{\Black{$*$}}}
    \Text(118,6)[lb]{\LARGE{\Black{$*$}}}
    \SetWidth{0.1}
    \Bezier(144,32)(144,0)(64,0)(64,32)\Line[arrow,arrowpos=0.165,arrowlength=3.5,arrowwidth=1.4,arrowinset=0.2,flip](138.511,19.098)(138.185,18.774)%JaxoID:FBez[arrow,arrowpos=0.165,arrowlength=3.5,arrowwidth=1.4,arrowinset=0.2,flip]
    \Text(324,16)[lb]{\normalsize{\Black{$d$}}}
    \Text(345,9)[lb]{\LARGE{\Black{$*$}}}
    \Text(413,26)[lb]{\LARGE{\Black{$*$}}}
    \SetWidth{0.8}
    \Photon(376,56)(376,8){5}{3}
    \Text(144,14)[lb]{\normalsize{\Black{$u$}}}
    \Text(332,-4)[lb]{\small{$(\Red{-})$}}
    \Text(446,28)[lb]{\normalsize{\Black{$K^0$}}}
    \Photon(541,55)(541,10){5}{3}
    \Text(520,-48)[lb]{\normalsize{\Black{$(c)$}}}
    \Text(476,14)[lb]{\normalsize{\Black{$d$}}}
    \Text(514,62)[lb]{\normalsize{\Black{$s$}}}
    \Text(558,-2)[lb]{\normalsize{\Black{$c$}}}
    \Text(536,-8)[lb]{\small{$(\Red{-})$}}
    \Text(525,-6)[lb]{\normalsize{\Black{$d$}}}
    \Text(504,-2)[lb]{\normalsize{\Black{$u$}}}
    \Text(498,9)[lb]{\LARGE{\Black{$*$}}}
    \Text(514,4)[lb]{\LARGE{\Black{$*$}}}
    \Text(508,26)[lb]{\normalsize{$(\Red{-})$}}
    \Text(498,-16)[lb]{\normalsize{\Black{$(d)$}}}
    \Text(548,26)[lb]{\normalsize{\Black{$W^+$}}}
    \Text(558,60)[lb]{\normalsize{\Black{$c$}}}
  \end{picture}
}
\end{center}
\caption{a) $\mathcal{O}\left(N_c^{1/2}\right)$, b) $\mathcal{O}\left(N_c^{-1/2}\right)$ and c) charm penguin contributions to the $K^0\to\pi\pi$ amplitudes. Stars denote again projections on the $\pi$ final states. Compared with Fig.\ref{Fig:Kp}, the extra minus signs result from the $d\bar{d}$ component of the $\pi^0$ in (b) and from the $V_{cd}$ CKM matrix element in (c), leading thus to an overall $\Delta I=1/2$ enhancement.}
\label{Fig:K0}
\end{figure}
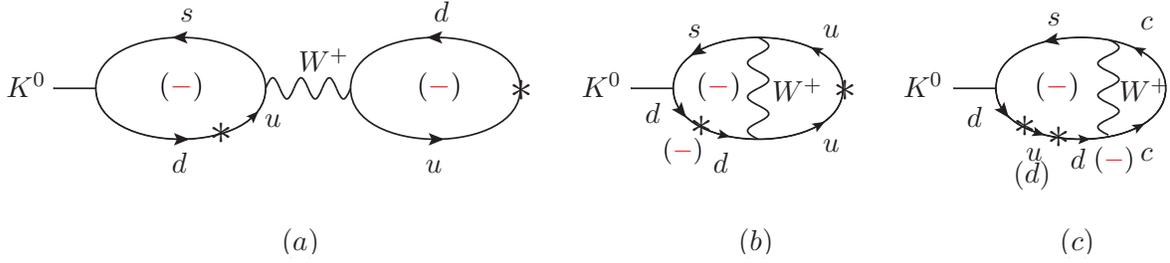

\section{A phenomenological estimate of $\hat{B}_K$}

A value of $\hat{B}_K$ below $3/4$ is also favored by the following  phenomenological approach based on  the observed $K^+ \to \pi^+ \pi^0$ weak decay.   We assume the absence of important new physics contributions in this tree-level process.        Applying then the standard operator product expansion and neglecting again isospin breaking effects, the corresponding amplitude reads:
\begin{equation}
\label{R17}
A(K^+ \to \pi^+\pi^0) = \left(\frac{G_F}{\sqrt{2}}\right) V_{ud} V_{us}^\ast \langle \pi^+ \pi^0 | \{ c_1 Q_1 + c_2 Q_2\} | K^+ \rangle
\end{equation}
with
\begin{eqnarray}
\label{R18}
Q_1 = \Bigl[\bar{s} \gamma_\mu (1-\gamma^5)d\Bigr] \otimes \Bigl[\bar{u} \gamma^\mu (1-\gamma^5)u\Bigr]
\nonumber\\
Q_2 =  \Bigl[\bar{s} \gamma_\mu (1-\gamma^5)u\Bigr] \otimes \Bigl[\bar{u} \gamma^\mu (1-\gamma^5)d\Bigr] 
\end{eqnarray}
the (non-penguin) $|\Delta S| = 1$ current-current operators.   Inspired  by Eq.(\ref{R13}), let us also normalize their hadronic matrix elements with respect to the VIA \footnote{In general,  $\langle \pi^+ | \bar{s} \gamma_\mu (1-\gamma^5) d | K^+  \rangle \equiv q^+_\mu f_+ (q^2_-) + q^-_\mu f_- (q^2_-)$  with  $q^\pm \equiv p_K \pm p_\pi$ and $f_+(0)=1$, $f_-(0)= f_K/f_\pi-1 \approx 0.2$  at first order in $SU(3)$ breaking. Here, we simply take $f_+ = 1$ and $f_- = 0$, having in mind that lattice QCD data give $f_+(0) = 0.956(8)$ \cite{Colangelo:2010et}.}: 
\begin{equation}
\label{R19}
\langle \pi^+ \pi^0 | Q_{1,2} | K^+ \rangle \equiv \left(1+\frac{1}{N_c}\right) B_K \left(\frac{-if_\pi}{\sqrt{2}}\right) (M^2_K - M^2_\pi). 
\end{equation}
In the leading logarithmic approximation, the Wilson coefficient associated with the relevant $\Delta I =  3/2$ operator in Eq.(\ref{R17}) is given by           
\begin{equation}
\label{R20}
 c_+ (\mu) \equiv c_1 + c_2 = [\alpha_s (M_W)/\alpha_s(\mu)]^a, \quad a = (9- 9/N_c)/(11N_c-2n_f).
\end{equation}
Consequently, we can express the renormalization scale independent $\hat{B}_K$ parameter in terms of the measured   $K^+ \to \pi^+ \pi^0$ decay amplitude:    
\begin{eqnarray}
\label{R21}
\hat{B}_K &\equiv& B_K (\mu) [\alpha_s(\mu)]^{-a}
\nonumber\\
&=& (\frac{3}{2}) [\alpha_s (M_W)]^{-a} [G_F \sin \theta_c \cos \theta_c f_\pi (M^2_K - M^2_\pi)]^{-1} | A(K^+ \to \pi^+ \pi^0)|.
\end{eqnarray}
Neglecting threshold effects and performing the full evolution from $M_W$ to $\mu$ in an effective four flavor theory, namely $n_f = 4$,  we obtain then
\begin{equation}
\label{R25}
\hat{B}_K (\mbox{pheno}) \simeq 0.6.
\end{equation}
 This value is not so far away from the lattice calculations (\ref{R3}) but below the large-$N_c$ prediction (\ref{R4}), if (at most) a 20\% theoretical uncertainty can be conceded for Eq.(\ref{R25}). This might be allowed in  such a rather phenomenological approach mainly relying on \textit{measured} form factors.       Indeed, the physical $f_{K,\pi}$ decay constants are consistently factorized with respect to the VIA, while  the $B$-parameter for the $\Delta I =  3/2$ operator introduced in Eq.(\ref{R19})   is assumed to be the same as the $B$-parameter for the $|\Delta S| = 2$ one defined in Eq.(\ref{R13}).   This  only assumption  sounds quite reasonable since the two operators have identical  QCD corrections above one GeV and Fierz relations below one GeV, in their evolution down to the hadronic scale  $\mu$. Moreover, it  can be fully justified at ${\cal O} (p^2)$ in a momentum expansion and in the flavor $SU(3)$ limit. In these limits indeed, both dimension-six current-current  operators transform like the symmetric and traceless irreducible representation (27$_L$,1$_R$) of the $SU(3)_L \times SU(3)_R$ chiral symmetry.   They belong thus to the single operator
\begin{eqnarray}
\label{R26}
O^{27} \propto \{ (\partial_\mu UU^\dagger)^j_i \otimes (\partial^\mu UU^\dagger)^\ell_k  &+& \left(\frac{1}{5}\right) [\delta^j_k \otimes (\partial_\mu U \partial^\mu U^\dagger)^\ell_i + \delta^\ell_i  \otimes (\partial_\mu U \partial^\mu U^\dagger)^j_k]
\nonumber
\\
&-&  \left(\frac{1}{20} \right) (\delta_k^j   \otimes \delta_i^\ell) (\partial_\mu U \partial^\mu U^\dagger)^m_m + (j \longleftrightarrow \ell)  \}
\end{eqnarray}
with
\begin{equation}
\label{R27}
U(x) \equiv \exp (i\frac{ \sqrt{2} \lambda_\alpha \pi^\alpha}{f}),
\end{equation}
the unitary Goldstone boson field transforming as $(3_L, \bar{3}_R)$ under this chiral group. The dimen\-sion-six effective operator $O^{27}$ is consistent with the standard normalization in Eq.(\ref{R13}) if proportional to $B_K f_K^2 f^2$.    It is also consistent with our normalization in Eq.(\ref{R19}) if, in addition, the scale factor $f$ is taken equal to $f_K^2/f_\pi$ with \cite{Colangelo:2010et}
\begin{equation}
\label{R28}
\frac{f_K}{f_\pi} \simeq 1.20.
\end{equation}
This flavor $SU(3)$ link between Eq.(\ref{R13}) and Eq.(\ref{R19}) not only sustains Eq.(\ref{R25}) within 20\%, but also explains a puzzling prediction on $\hat{B}_K$ at leading order in the chiral perturbation theory.    For strong interactions at low momenta, the scale factor $f$ introduced in Eq.(\ref{R27}) is naturally identified with the pion decay constant $f_\pi$, as it has been done implicitly in \cite{Donoghue:1982cq}. Consequently, in this well-defined theoretical frame we have to substitute $f^2_K/f_\pi$ for $f_\pi$ in Eq.(\ref{R19}).     The predicted value for $\hat{B}_K$ is thus reduced by about 30\%, 
\begin{equation}
\label{R31}
\hat{B}_K (\mbox{chiral}) = \Bigl(\frac{f_\pi}{f_K}\Bigr)^2 \hat{B}_K (\mbox{pheno}) \simeq 0.4,                             
\end{equation}
 in very good agreement with next-to-leading order in $1/N_c$ calculations of $\hat{B}_K$ implemented in the strict chiral limit, i.e., with massless quarks \cite{Bijnens:2006mr}. So, unlike the phenomenological approach developed here-above, the chiral one appears to be extremely sensitive to the $SU(3)$-splitting (\ref{R28}) observed in the weak decay constants. This high sensitivity on the SU(3) breaking   $(m_d \neq m_s)$ also reflects itself at the ${\cal O}(p^4)$ in the chiral perturbation theory for the $|\Delta S| = 2$ and the $\Delta I =  3/2$ matrix elements: the corresponding one-loop logarithmic corrections proportional to $M^4_K \ln (M^2_K/\mu^2_\chi)$ are quite different from each other.  In particular, these one-loop corrections turn out to be sizeable for $B_K$ and perturbation theory even breaks down if the chiral scale $\mu_\chi$ is taken at one GeV \cite{Bijnens:1984ec}. However, at this point, it is worth recalling again the $1/N_c$ matching approach based on a duality between QCD and a meson theory for strong interactions.     First introduced in \cite{Bardeen:1987vg} to get rid of the unknown ${\cal O}(p^4)$ corrections arising from effective operators with four derivatives, this approach requires a lower chiral scale, i.e., $\mu_\chi$ around (0.6 $-$ 0.7) GeV \cite{Gerard:1990dx}.

\section{Conclusion}

Three very well measured CP-asymmetries are currently at our disposal in Eqs.(\ref{K2a}-\ref{K2c}). So, they should in principle provide a double check of the Standard Model with a single CKM phase. Unfortunately, large hadronic uncertainties on the $|\Delta S| = 1$ penguin operators involved in both $\varepsilon_K$ and $\varepsilon'_K$ prevent us from doing this so far. In particular, the information contained in the Kaon direct CP-asymmetry (\ref{K2a}) cannot be really exploited yet. 
\\
The main point of our work is to show that such is not the case for the pure  $|\Delta S| = 2 \  K^0 \leftrightarrow \bar K^0$ transition at the origin of the Kaon indirect CP-asymmetry (\ref{K2c}). Indeed, a $1/N_c$ expansion within the Standard Model gives as a result
\begin{equation}
\label{R30}
2 \ \mbox{Re}(\varepsilon_K)   = [\hat{B}_K / 0.75] (2.96 \pm 0.38) \ 10^{-3}
\end{equation}
at the next-to-leading order, once the CP-asymmetry (\ref{K2b}) is used as an input. Moreover, we argue that the Kaon $B$-parameter is bounded from above by its large-$N_c$ value, namely
\begin{equation}
\label{R29}
\hat{B}_K   \leq 0.75,
\end{equation}
in a way consistent with both the empirical  $\Delta I = 1/2$ rule and phenomenological flavor SU(3). This upper bound on $\hat{B}_K$ turns out to be also (almost) saturated by the lattice QCD calculations. Consequently, the theoretical value of the Kaon CP-asymmetry appears to be only one sigma away from its experimental one. In fact, the 13\% uncertainty in Eq.\eqref{R30} is mainly due to the 3\% uncertainty on the $V_{\mbox{{\scriptsize cb}}}$ element, through the unitarity conditions on the CKM matrix. So, the Kaon indirect CP violation provides us with a rather interesting test of the Standard Model. For illustration, one may substitute the Re$(\varepsilon_K)$ bound for the $\varepsilon_K$ band in fits of the CKM unitary triangle.

\acknowledgments

We thank Andrzej Buras for his encouragements and very useful comments, and Elvira Cervero Garcia and Philippe Mertens for interesting discussions.\\ 
This work dedicated to the memory of Joaquim Prades  has been supported by the Belgian Federal Office for Scientific, Technical and Cultural Affairs through the Interuniversity Attraction Pole No P6/11.

\bibliographystyle{JHEP}
\bibliography{biblio.bib}

\end{document}